\documentclass[aps,prl,preprintnumbers,showpacs,twocolumn,nofootinbib]{revtex4}
\newcommand{\PRE}[1]{}       
\usepackage{graphicx}
\usepackage{dcolumn}
\usepackage{bm}
\def\etal{{\em et~al.}}

\def\Mpc{\text{Mpc}}

\def\vol#1  {{{#1}{\rm,}\ }}

\newcount\refno
\newcount\rfno
\def\eq{$^{\the\refno\ }$\advance\refno by 1}
\def\ad{\advance\rfno by 1}

\newcommand{\Gravitino}{\tilde{G}}

\newcommand{\sneu}{\tilde{\nu}}
\newcommand{\gev}{\text{GeV}}
\newcommand{\ev}{\text{eV}}
\newcommand{\s}{\text{s}}

\begin{document}

\preprint{UCI-TR-2005-27}  

\title{
\PRE{\vspace*{2.0in}}
SuperWIMP Solutions to Small Scale Structure Problems
\PRE{\vspace*{.4in}}
}

\author{Jose~A.~R.~Cembranos}
\author{Jonathan~L.~Feng}
\author{Arvind Rajaraman}
\author{Fumihiro Takayama
\PRE{\vspace{.1in}}
}

\affiliation{Department of Physics and Astronomy,
 University of California, Irvine, CA 92697, USA
\PRE{\vspace{.5in}}
}


\begin{abstract}
\PRE{\vspace*{.2in}} Collisionless, cold dark matter in the form of
weakly-interacting massive particles (WIMPs) is well-motivated in
particle physics, naturally yields the observed relic density, and
successfully explains structure formation on large scales.  On small
scales, however, it predicts too much power, leading to cuspy halos,
dense cores, and large numbers of subhalos, in apparent conflict with
observations. We consider superWIMP dark matter, produced with large
velocity in late decays at times $10^5~\s - 10^8~\s$.  As analyzed by
Kaplinghat in a more general setting, we find that superWIMPs have
sufficiently large free-streaming lengths and low phase space
densities to help resolve small scale structure problems while
preserving all of the above-mentioned WIMP virtues.
\end{abstract}



\maketitle


The microscopic identity of dark matter (DM) is one of the major
puzzles in basic science today.  In the current standard cosmological
picture, the Universe contains non-baryonic dark matter with abundance
$\Omega_{\text{DM}} h^2 = 0.095 - 0.129$~\cite{Spergel:2003cb}, where
$\Omega_{\text{DM}}$ is the energy density in units of the critical
density, and $h \simeq 0.71$ is the reduced Hubble parameter.  This
component is typically assumed to be cold, collisionless, and
non-self-interacting dark matter, which we refer to as CDM throughout
this work.  CDM is remarkably successful in explaining the observed
large scale structure down to length scales of $\sim 1~\Mpc$.

Among the most well-motivated CDM candidates are weakly-interacting
massive particles (WIMPs), with masses of the order of the weak scale
$M_W \sim 100~\gev$ and interaction cross sections $\sigma \sim g^2
M_W^{-2}$.  WIMPs emerge naturally from several well-motivated
particle physics frameworks and include the lightest supersymmetric
particle (LSP) in $R$-parity conserving supersymmetry
models~\cite{SUSY}, the lightest Kaluza-Klein state in models with
universal extra dimensions~\cite{UED}, and branons in brane-world
models~\cite{BW}.  In addition, WIMPs naturally have thermal relic
densities of the desired order of magnitude.

Despite its considerable successes, however, CDM appears to face
difficulty in explaining the observed structure on length scales $\alt
1~\Mpc$. Numerical simulations assuming CDM predict overdense cores in
galactic halos~\cite{cores}, too many dwarf galaxies in the Local
Group~\cite{dwarfs}, and have trouble producing enough disk galaxies
without angular momentum loss~\cite{Navarro:1999fr,cores}. Although
there is not currently consensus that the small scale problems of CDM
are insurmountable~\cite{CDMOK}, the number and variety of problems
put considerable pressure on CDM and have motivated many alternative
dark matter candidates.  These include self-interacting dark
matter~\cite{Spergel:1999mh}, collisional dark matter~\cite{coll},
thermal warm dark matter (WDM)~\cite{WDM}, annihilating dark
matter~\cite{Kaplinghat:2000vt}, non-thermal WIMP
production~\cite{Lin:2000qq}, and other proposals, such as the
possibility of a broken scale invariance in the power
spectrum~\cite{Kamionkowski:1999vp}.

A common feature of these new hypotheses is that they preserve the
successes of standard CDM on large scales, but reduce power on small
scales. Unfortunately, this virtue is achieved at a cost: in contrast
to WIMPs, these candidates are generally not well-motivated
independently by particle physics, and their relic abundance is also
not naturally in the correct range.  For example, to explain the
observed small scale structure, thermal WDM particles must have a mass
greater than about 500 eV~\cite{Viel:2005qj,Narayanan:2000tp}.  On the
other hand, the observed relic density is naturally achieved for
masses $\sim 10~\ev$.  To resolve this discrepancy requires either an
unreasonably large number ($\sim 10^{3}$) of light degrees of freedom
at the time of decoupling or, alternatively, a nonstandard cosmology
with a large injection of entropy at late times.

Here we consider superweakly-interacting massive particle (superWIMP)
DM.  In superWIMP scenarios, a WIMP freezes out as usual, but then
decays to a stable DM particle that interacts {\em
superweakly}~\cite{Feng:2003xh,Feng:2003uy}.  Examples of superWIMPs
include non-thermally produced weak-scale
gravitinos~\cite{Feng:2003xh,Feng:2003uy,Ellis:2003dn,Feng:2004zu,%
Roszkowski:2004jd}, axinos~\cite{axinos}, and
quintessinos~\cite{Bi:2003qa} in supersymmetry, and Kaluza-Klein
graviton and axion states in models with universal extra
dimensions~\cite{Feng:2003nr}.  SuperWIMPs preserve WIMP virtues: they
exist in the same well-motivated frameworks and naturally have the
right relic density, since they inherit it from late-decaying WIMPs.
This latter property and the effect on small scale structure discussed
here are absent for thermally-produced gravitinos.

In contrast to WIMPs, superWIMPs are produced with large velocities at
late times.  For example, gravitino or Kaluza-Klein graviton
superWIMPs are naturally produced at $\tau_X \sim
M_{\text{Pl}}^2/M_W^3 \sim 10^5~\s - 10^8~\s$, where the reduced
Planck mass $M_{\text{Pl}} \equiv (8\pi G_N)^{-1/2} \simeq 2.4 \times
10^{18}~\gev$ enters because these superWIMPs interact only
gravitationally.  This has two effects.  First, the velocity
dispersion reduces the phase space density, smoothing out cusps in DM
halos.  Second, such particles damp the linear power spectrum,
reducing power on small scales and improving consistency with
structure formation.  As we will show, these effects are sufficiently
strong that superWIMPs may provide a natural resolution to small scale
structure problems.  Similar conclusions have been reached in the more
general setting explored in Ref.~\cite{Kaplinghat}.

We first consider effects coming from the velocity dispersion. Such
effects may be characterized by $Q \equiv \rho/\langle
v^2\rangle^{3/2}$, the dark matter mass per unit volume of
6-dimensional phase space, where $\rho$ is the mass density and
$\langle v^2\rangle$ is the velocity
dispersion~\cite{Hogan:2000bv}. $Q$ has a number of important
properties.  The fine-grained value of $Q$ remains constant for
collisionless, dissipationless gases.  It may be determined
analytically in scenarios with decaying dark matter.  In addition, the
coarse-grained value of $Q$ can only decrease, a property that follows
from the relation of $Q$ to thermodynamic entropy.

The coarse-grained value of $Q$ can be estimated from rotation curves,
gas emission, and gravitational
lensing~\cite{Hogan:2000bv,Dalcanton:2000hn}.  Galaxies with
coarse-grained $Q$ near $Q_0 \equiv 1.0 \times 10^{-27}~\gev^4 \simeq
1.2 \times 10^{-4} \, (M_{\odot}/\text{pc}^3) / (\text{km}/\s)^3$ have
been observed~\cite{Dalcanton:2000hn}. Given the properties of $Q$
noted above, this imposes a lower limit on fine-grained $Q$ of $Q >
Q_0$. At the same time, it has been
argued~\cite{Hogan:2000bv,Dalcanton:2000hn} that values of $Q$ close
to $Q_0$ with, for example, $Q_0 \alt Q \alt 4 Q_0$, are preferred, as
they reduce the maximal cuspiness of galactic halos to that actually
observed.

We now determine the fine-grained value of $Q$ in superWIMP scenarios.
Throughout this work, we assume that superWIMPs are produced between
$10^{3}~\s$ and $10^{12}~\s$ so that the Universe is
radiation-dominated and the number of effective relativistic degrees
of freedom is constant.  We define a relativistic version of $Q$,
$\tilde{Q} \equiv \rho / \langle u^2 \rangle ^{3/2}$, where $u = p/m$
is the three-momentum normalized by the particle's mass.  Because $u$
redshifts, $\tilde{Q}$ has the advantage that it is invariant given
expansion in the Universe for both relativistic and non-relativistic
matter, but reduces to $Q$ at late times when the matter becomes
non-relativistic.

We may estimate the value of $Q$ at structure formation by determining
the value of $\tilde{Q}$ when superWIMPs are produced.  For
simplicity, we assume that all superWIMPs are produced at the decay
lifetime $\tau_X$.  The exponential distribution of production times
gives ${\cal O}(1)$ corrections to the results described here and will
be discussed in detail elsewhere~\cite{Kaplinghat,next}.  We find
\begin{equation}
\label{Q} Q = \tilde{Q}(\tau_X) \simeq Q_0\, u_X^{-3} 
\left[ \frac{10^6~\s}{\tau_X}\right]^{3\over2} 
\left[ \frac{\Omega_{\text{SWIMP}}h^2}{0.11} \right] \ ,
\end{equation}
where $u_X \equiv u(\tau_X)$.  Note that for $u_X \sim 1$ and $\tau_X
\sim 10^6~\s$, natural values in the cases of gravitino and
Kaluza-Klein graviton superWIMPs, the phase space density is in the
preferred range to eliminate cuspy halos.

We now turn to the effect on the power spectrum.  Initially, the
matter density has small inhomogeneities.  These inhomogeneities
evolve linearly at first, but eventually evolve non-linearly to form
the structure observed today.  In the present Universe, the scale of
non-linearity is expected to be around 30 Mpc.  After this point,
$N$-body simulations are necessary to analyze the evolution of the
power spectrum. These analyses show that CDM predicts an excess of
power on scales under 1 Mpc.

One way to reduce power on small scales and ameliorate this problem is
through the free-streaming of DM.  From its production time $\tau_X$
until matter-radiation equality at $t_{\text{EQ}}\simeq 2.2\times
10^{12}$ s, superWIMPs can stream out of overdense regions into
underdense regions, smoothing out inhomogeneities. A free-streaming
scale much larger than 1 Mpc is excluded by observations of Lyman
alpha clouds~\cite{Croft:2000hs}.  (For instance, for WDM, the
constraints $m_{\text{WDM}}\agt 550~\ev$~\cite{Viel:2005qj} and
$m_{\text{WDM}}\agt 750~\ev$~\cite{Narayanan:2000tp} correspond, given
our definition of free-streaming scale, to $\lambda_{\text{FS}}\alt
1.4~\Mpc$ and $\lambda_{\text{FS}}\alt 1.0~\Mpc$, respectively.)
However, values close to this (roughly, $1.0~\Mpc \agt
\lambda_{\text{FS}} \agt 0.4~\Mpc$) could resolve the present
disagreements.

The free-streaming scale for superWIMP dark matter can be estimated to
be
\begin{eqnarray}\label{3lold}
\lambda_{\text{FS}} = 
\int^{t_{\text{EQ}}}_{\tau_X}\frac{v(t)dt}{a(t)} =
\lambda(t_{EQ})-\lambda(\tau_{X})\,,
\end{eqnarray}
where $a(t)$ is the cosmic scale factor, $v(t)$ is the
superWIMP velocity, and 
\begin{eqnarray}
\lambda(t)&=&2R_{\text{EQ}}\,u_{\text{EQ}}\;
\ln \left[ \frac{1}{u(t)}+\sqrt{1+\frac{1}{u^2(t)}}\ \right] 
\ ,
\label{lQ}
\end{eqnarray}
where $R_{\text{EQ}}\equiv c\,t_{\text{EQ}}/a(t_{\text{EQ}})\simeq
93~\Mpc$ and $u_{\text{EQ}}\equiv
u(t_{\text{EQ}})$~\cite{Lin:2000qq,Bi:2003qa}. In the common case that
superWIMPs are relativistic when produced ($u_{X}\agt 1$) but
non-relativistic at $t_{\text{EQ}}$ ($u_{\text{EQ}}\ll 1$),
Eq.~(\ref{3lold}) simplifies to
\begin{eqnarray}
\lambda_{\text{FS}} &\approx& \lambda(t_{EQ})\approx
1.0~\Mpc
\ u_X \left[\frac{\tau_X}{10^6~\s}\right]^{1\over2}
\\
&&\times \left\{ 1+0.14
\ln \left[ \left(\frac{10^6~\s}{\tau_X}\right)^{1\over2}
\frac{1}{u_X} \right] \right\} \ , \nonumber
\end{eqnarray}
demonstrating that production times of $\sim 10^6~\s$ naturally also
provide the preferred order of magnitude for the free-streaming scale.
Note that in this case, $Q$ and $\lambda_{\text{FS}}$ are both
functions of $u_X \tau_X^{1/2}$, and so are correlated, as they are
for WDM.  However, in the superWIMP scenario with $u_X \alt 1$,
$\lambda_{\text{FS}} < \lambda(t_{EQ})$ and this degeneracy is broken,
opening up new possibilities~\cite{next}.

\begin{figure}[bt]
\centerline{
\includegraphics[
width=7.5cm,height=7.5cm,clip]{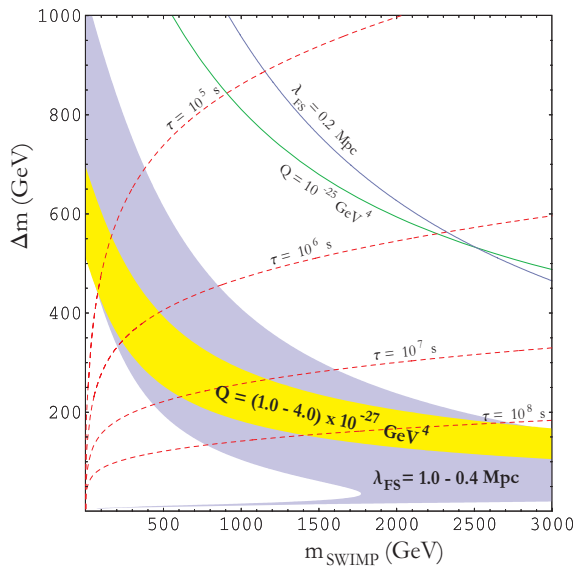}} 
\caption{Preferred regions (shaded) of phase space density $Q$ and
free-streaming length $\lambda_{\text{FS}}$ in the $(m_{\text{SWIMP}},
\Delta m)$ plane, where $\Delta m \equiv m_{\text{NLSP}} -
m_{\text{SWIMP}}$, for gravitino superWIMPs with a sneutrino NLSP.
The regions under both bands are disfavored.  In the regions above
both bands, superWIMP dark matter becomes similar to CDM;
representative values of $Q$ and $\lambda_{\text{FS}}$ are shown.
Contours of typical lifetimes $\tau_{\tilde{\nu}}$ are also shown.  We
have assumed $\Omega_{\text{SWIMP}}h^2=0.11$. }
\label{sneu}
\end{figure}

We now consider the particular case of gravitino superWIMPs.  The
next-to-lightest supersymmetric particle (NLSP) may be either a scalar
(e.g., a sneutrino or a slepton) or a fermion (a neutralino).  If the
NLSP is a sneutrino, gravitinos are produced at
time~\cite{Feng:2003uy}
\begin{equation}
 \tau_{\sneu}
 =48\pi M_{\text{Pl}}^2\frac{m_{\Gravitino}^2}{m_{\sneu}^5}
\left[1 - \frac{m_{\Gravitino}^2}{m_{\sneu}^2} \right]^{-4}
\label{sneulifetime}
\end{equation}
with $u_X = (m_{\sneu}^2 - m_{\Gravitino}^2) / (2 m_{\sneu}
m_{\Gravitino})$. These then determine $Q$ and $\lambda_{\text{FS}}$
through Eqs.~(\ref{Q})--(\ref{lQ}); the results are shown in
Fig.~\ref{sneu}.  Remarkably, preferred values of $Q$ and
$\lambda_{\text{FS}}$ are simultaneously realized in superWIMP
scenarios with natural, weak-scale masses with $m_{\text{NLSP}} \agt
500~\gev$.  The superWIMP scenario is also constrained by Big Bang
nucleosynthesis and the Planckian spectrum of the cosmic microwave
background.  These constraints~\cite{Jedamzik:2004er} have been
evaluated in several
studies~\cite{Feng:2003xh,Feng:2003uy,Ellis:2003dn,%
Feng:2004zu,Roszkowski:2004jd,Hu:1993gc}.  For a $\tilde{\nu}$ NLSP, the
main constraint comes from hadronic and electromagnetic energy
produced in the three-body decays $\tilde{\nu} \to l W \tilde{G}, \nu
Z \tilde{G}$.  These modes have small branching ratios of the order of
$10^{-3}$ and so the parameter space is not strongly constrained for
$\tau_X \agt 10^6~\s$~\cite{Feng:2004zu}.

For a charged slepton NLSP, the lifetime is identical to that given in
Eq.~(\ref{sneulifetime}).  However, charged sleptons are not
collisionless and are electromagnetically coupled to the baryon-photon
plasma.  The opposite tendencies of pressure repulsion and
gravitational attraction generate acoustic waves with density
perturbation oscillations of photons, baryons and sleptons. After
gravitino production, however, the photon-baryon fluid is coupled only
gravitationally to the neutral superWIMP.  Power is therefore reduced
on scales that enter the horizon before this decoupling, and this
effect can be more important than the free-streaming damping discussed
above~\cite{Sigurdson:2003vy,next}.

\begin{figure}[bt]
 \centerline{
    \includegraphics[width=7.5cm,height=7.5cm,clip]{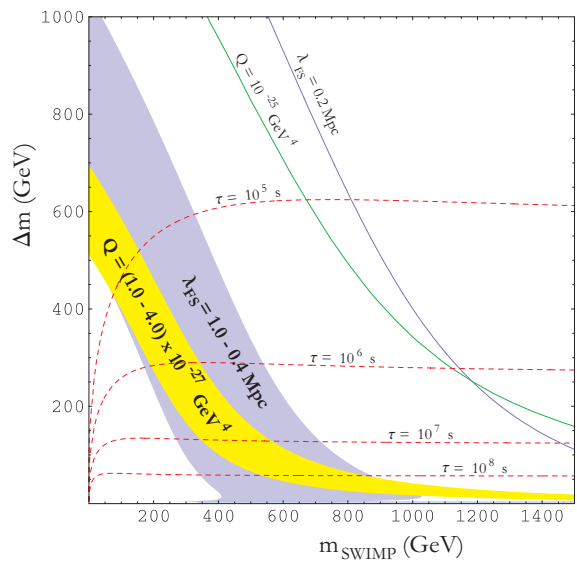}}
  \caption{Same as in Fig.~\protect\ref{sneu}, but for gravitino
superWIMPs with a photino NLSP.}  \label{ph}
\end{figure}

If the decaying WIMP is a neutralino, the superWIMP production time is
different. In general, the neutralino is a mixture of neutral Bino,
Wino, and Higgsino states.  For the specific case of a photino, the
lifetime is~\cite{Feng:2003uy}
\begin{equation}
\tau_{\tilde{\gamma}}
=  48\pi M_{\text{Pl}}^2 \frac{m_{\Gravitino}^2} {m_{\tilde{\gamma}}^5}
\left[1 - \frac{m_{\Gravitino}^2}{m_{\tilde{\gamma}}^2} \right]^{-3}
\left[1 + 3 \frac{m_{\Gravitino}^2}{m_{\tilde{\gamma}}^2}
\right]^{-1} \,. \label{photinolifetime}
\end{equation}
The resulting $Q$- and $\lambda_{FS}$-preferred regions are shown in
Fig.~\ref{ph}.  As in the sneutrino NLSP case, preferred values for
both quantities are obtained for natural weak-scale masses; if
anything, the preferred superWIMP and NLSP masses are lighter and more
natural.  For typical neutralinos, hadrons are produced in two-body
decays, leading to extremely severe constraints from Big Bang
nucleosynthesis.  For photinos, however, the constraints are relaxed,
since hadrons are produced only in three-body decays, and the
superWIMP resolution to small scale structure problems may be
realized.

In conclusion, we have examined the implications of superWIMP dark
matter for small scale structure.  Because superWIMP dark matter is
produced with large velocity in late decays, its phase space density
is decreased, smoothing out halo cusps.  At the same time, superWIMPs
damp the linear power spectrum, which lowers the concentration of dark
matter in the cores of galactic halos.  This effect may bring
numerical simulations into agreement with observations of dwarf
galaxies and reduce the excess of dwarf galaxies relative to CDM
predictions.

The effects on small scale structure may also be achieved by WDM or
dark matter with exotic interactions.  In contrast to those
possibilities, however, superWIMPs are automatically present in
particle physics models with supersymmetry or extra dimensions and are
naturally produced with the correct relic density.  Dark matter
produced in late decays will necessarily be warmer than CDM.  It is
remarkable, however, that for superWIMP gravitinos, where the
production times and velocities are determined by the fixed energy
scales $M_{\text{EW}}$ and $M_{\text{Pl}}$, the predicted values of
both $Q$ and $\lambda_{\text{FS}}$ are in favorable ranges to resolve
outstanding problems without violating other constraints from
cosmology and particle physics.  SuperWIMPs therefore appears to
combine the most appealing features of both CDM and WDM.  This
explanation will be probed by future observations, especially those
constraining the epoch of reionization~\cite{Kaplinghat,Reion}:
reionization by redshift 6 implies $Q \agt 0.1\times Q_0$, compatible
with the preferred values analyzed in this work, but confirmation of
indications from WMAP of earlier reionization could restrict parameter
space greatly. At the same time, given the virtues described here, it
would be especially interesting to see if the promise of superWIMPs is
realized by $N$-body simulations of structure formation and
semi-analytic analyses.

{\em Acknowledgments} --- We are grateful to J.~Bullock and
M.~Kaplinghat for valuable conversations.  The work of JARC is
supported in part by NSF grant No.~PHY--0239817, the Fulbright-MEC
program, and the BFM 2002-01003 project (DGICYT, Spain). The work of
JLF is supported in part by NSF grant No.~PHY--0239817, NASA Grant
No.~NNG05GG44G, and the Alfred P.~Sloan Foundation. The work of AR is
supported in part by NSF Grant No.~PHY--0354993.  The work of FT is
supported in part by NSF grant No.~PHY--0239817.

\end{document}